\documentclass[aps,prb,amsmath,amssymb,preprint,showpacs,showkeys,superscriptaddress]{revtex4-1}

\usepackage{graphicx}
\usepackage{dcolumn}
\usepackage{color}
\usepackage{ulem}

\begin{document}

\title{Excitation intensity dependence of photoluminescence spectra of SiGe quantum dots grown on prepatterned Si substrates: evidence for biexcitonic transition}

\author{P. Klenovsk\'y}\email{klenovsky@physics.muni.cz}\affiliation{Department of Condensed Matter Physics, Faculty of Science, Masaryk University, Kotl\'a\v{r}sk\'a~2, 61137~Brno, Czech~Republic},\affiliation{CEITEC - Central European Institute of Technology, Masaryk University, Kamenice 753/5, 62500~Brno, Czech~Republic}
\author{M. Brehm}\affiliation{Institute of Semiconductor and Solid State Physics, Johannes Kepler University Linz, 4040~Linz, Austria}
\author{V. K\v{r}\'apek}\affiliation{Department of Condensed Matter Physics, Faculty of Science, Masaryk University, Kotl\'a\v{r}sk\'a~2, 61137~Brno, Czech~Republic},\affiliation{Institute of Physics, Academy of Sciences of the Czech Republic, Cukrovarnick\'a 10, Praha 6, 162 53, Czech~Republic}
\author{E. Lausecker} \affiliation{Institute of Semiconductor and Solid State Physics, Johannes Kepler University Linz, 4040~Linz, Austria}
\author{D. Munzar}\affiliation{Department of Condensed Matter Physics, Faculty of Science, Masaryk University, Kotl\'a\v{r}sk\'a~2, 61137~Brno, Czech~Republic},\affiliation{CEITEC - Central European Institute of Technology, Masaryk University, Kamenice 753/5, 62500~Brno, Czech~Republic}
\author{F. Hackl}\affiliation{Institute of Semiconductor and Solid State Physics, Johannes Kepler University Linz, 4040~Linz, Austria}
\author{H. Steiner}\affiliation{Institute of Semiconductor and Solid State Physics, Johannes Kepler University Linz, 4040~Linz, Austria}
\author{T. Fromherz}\affiliation{Institute of Semiconductor and Solid State Physics, Johannes Kepler University Linz, 4040~Linz, Austria}
\author{G. Bauer}\affiliation{Institute of Semiconductor and Solid State Physics, Johannes Kepler University Linz, 4040~Linz, Austria}
\author{J. Huml\'i\v{c}ek}\affiliation{Department of Condensed Matter Physics, Faculty of Science, Masaryk University, Kotl\'a\v{r}sk\'a~2, 61137~Brno, Czech~Republic},\affiliation{CEITEC - Central European Institute of Technology, Masaryk University, Kamenice 753/5, 62500~Brno, Czech~Republic}

\date{\today}

\begin{abstract}

The pumping intensity ($I$) dependence of the photoluminescence (PL) spectra of perfectly laterally two-dimensionally ordered SiGe quantum dots on Si(001) substrates was studied. The PL results from recombinations of holes localized in the SiGe quantum dots and electrons localized due to the strain field in the surrounding Si matrix. The analysis of the spectra revealed several distinct bands, attributed to phonon-assisted recombination and no-phonon recombination of the excitonic ground state and of the excited excitonic states, which all exhibit a linear $I$ dependence of the PL intensity. At approximately $I$\,$\geq$\,3\,Wcm$^{-2}$, additional bands with a nearly quadratic $I$ dependence appear in the PL spectra, resulting from biexcitonic transitions. These emerging PL contributions shift the composite no-phonon PL band of the SiGe quantum dots to higher energies. The experimentally obtained energies of the no-phonon transitions are in good agreement with the exciton and biexciton energies calculated using the envelope function approximation and the configuration interaction method.

\end{abstract}



\maketitle

\section{Introduction\label{intro}}

SiGe quantum dots (QDs) are of interest, since they provide a promising way towards an infra-red light source operating on telecom wavelengths, integrable into the present Si-based technology.~\cite{Tsybeskov} 
Although these structures have been studied for quite a long time,~\cite{Mo, Eaglesham} there are only a few reports on the excitonic structure of their photoluminescence (PL) spectra.~\cite{Yakimov1,Yakimov,Rastelli} The large size of the QDs and an inhomogeneous Ge distribution due to intermixing and 
clustering lead to a pronounced broadening of the PL lines of QD ensembles.~\cite{BrehmNRL} Additional broadening arises from the indirect nature of the optical transitions both in real and reciprocal space. 
Due to the broad spectra, an unambiguous assignment of the electronic QD transitions of randomly nucleated QDs grown on planar Si(001) has not been reported so far. For the same reasons, up to now only continuous linear and sub-linear shifts of the PL emission energies with increasing excitation intensity ($I$) could be observed both for SiGe QDs \cite{Wan,Dashiell} and quantum wells (QWs),~\cite{Thewalt,Penn1,Penn2} which were attributed to state filling effects and carrier induced band-bending as a consequence of the spatial separation of the excited electrons and holes.

It has been shown that by a precise positioning of the QDs on pre-structured Si substrates,~\cite{Zhong1} a significant narrowing of the PL spectra can be achieved,~\cite{BrehmNRL, Stoica, Lausec, Florian} making a more detailed analysis of the dependence of the PL spectra on the excitation intensity possible. 

However, no decomposition of the PL spectra in terms of individual excitonic recombinations has been reported so far for laterally ordered SiGe QDs.
Here we provide an identification of both bound excitonic and biexcitonic transitions in the PL spectra of ensembles of ordered SiGe QDs, using the linear and quadratic excitation intensity dependence of those transitions. This identification is facilitated by the clear observation of additional PL emission lines that appear with  increasing $I$ at higher photon energies as opposed to the continuous shift of the PL spectra reported up to now. 
The assignments are supported by the results of exciton and biexciton energy level calculations.

Extensive work on the exciton-exciton interaction energies has been reported for self-assembled epitaxial III-V compound quantum dots~\cite{SkolnickARoMR2004,NarvaezPRB2005,SchliwaPRB2009,MatsudaAPL2007,BansalPRB2009,Kle} as well as for chemically synthesized colloidal nanocrystals~\cite{DeutschPCCP2011,KambhampatiTJoPCC2011,KambhampatiTJoPCL2012} for material systems with both type-I and II band alignment. The understanding of the multi-exciton interaction and the exciton relaxation dynamics has been shown to be important for the realization of hot carrier extraction and optical gain.~\cite{KlimovN2007, KambhampatiTJoPCL2012}

In this work the exciton-exciton interaction as well as the exciton excited state energies are determined experimentally for SiGe QDs epitaxially embedded in a Si matrix. In this material system, the holes are confined in the SiGe QDs and the electrons to the tensile strained regions in the Si matrix surrounding them, forming a type-II system. Our results show that for the ordered, highly uniform SiGe QDs investigated in this work, the energy splitting between the biexciton and exciton emission line is larger than the line width, which is a prerequisite to achieve optical gain for single exciton states.~\cite{KlimovN2007}

This paper is organized as follows: in Sec.\,\ref{exp}, experimental details on SiGe island growth and on the PL measurements are presented. In Sec.\,\ref{res}, the analysis of the PL spectra, their excitation intensity dependence, electronic structure calculations and the resulting assignment of the exciton, excited exciton, biexciton and excited biexciton energies are given. In Sec.\,\ref{conc}, the conclusions are presented.                    

\section{Experimental \label{exp}}

\begin{figure}[ht]
\renewcommand{\tabcolsep}{2pt}
\begin{tabular}{c}
\includegraphics[width=0.4\textwidth]{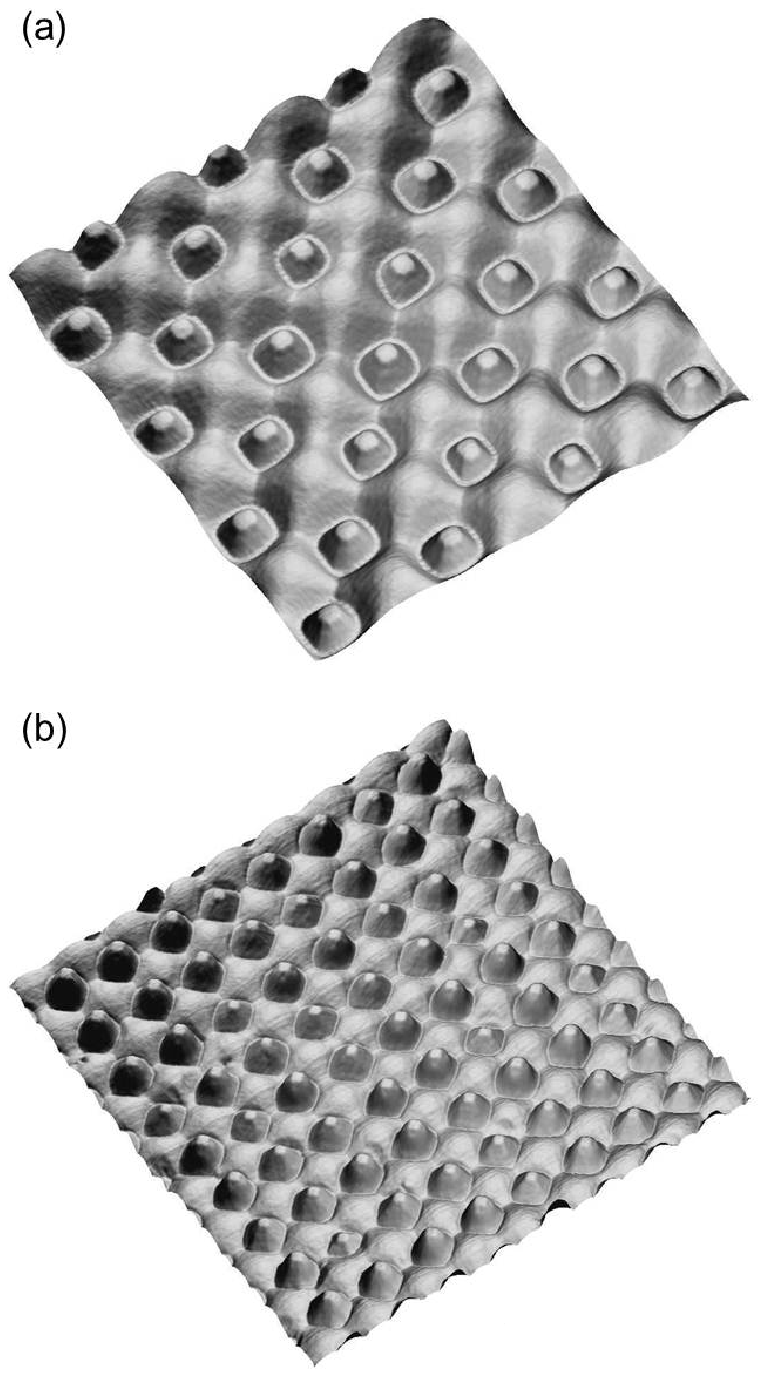} \\
\end{tabular}
\caption{
Atomic force micrographs of the samples (a) S1 (pit period 300~nm) and (b) S2 (pit period 170~nm). The scanned area was $1.5\times 1.5\,\mathrm{\mu m}$ for both samples. The average height of the QDs in  sample S1 (S2) is 15.3~(15)~nm and the QD base diameter is 122.8~(118)~nm.
\label{figAFM}}
\end{figure}

Two samples with different average Ge content in the SiGe QDs were investigated (S1 and S2). High resistivity p-type Si (001) substrates were pit patterned by nanoimprint lithography and subsequent pattern transfer into  the substrate
by reactive ion etching. 
The pattern periods were 300~nm for S1 and 170~nm for S2, and the patterned areas were 3$\times$3\,mm$^2$ for both samples. 
The resulting pits had diameters and depths of 160~nm and 47~nm for S1, and 120~nm and 35~nm for S2, respectively. 
After standard cleaning, the wafers were in-situ degassed in the solid source molecular beam epitaxy chamber 
for 40~min at $720\,\mathrm{^{\circ}C}$. Hereafter, a 45~nm thick Si buffer layer (growth rate $R=0.6\,\mathrm{\AA /s}$) was deposited at a substrate temperature that was increased from $450\,\mathrm{^{\circ}C}$ to $550\,\mathrm{^{\circ}C}$ followed by the deposition of 6 monolayers (ML) of Ge at $690\,\mathrm{^{\circ}C}$ ($R=0.05\,\mathrm{\AA /s}$) for S1 and 8.3 ML of Ge at $625\,\mathrm{^{\circ}C}$ 
($R=0.025\,\mathrm{\AA /s}$) for S2. Subsequently, a 50\,nm or a 10\,nm thick Si capping layer  was 
deposited at a temperature as low as $300\,\mathrm{^{\circ}C}$ in order to avoid unwanted Si incorporation and QD shape transformations for S1 or S2, respectively.~\cite{BrehmJAP} Atomic force micrographs of the ordered and capped islands  are shown in Fig.~\ref{figAFM} for S1 and S2.

In the PL measurements, performed at $4.2\,\mathrm{K}$, the samples were excited by an Ar$^+$ laser tuned to the wavelength of 457.9\,nm, focused by a lens to a circular area with 400~$\mathrm{\mu m}$ diameter ($\sim 10^{6}$ irradiated QDs). The excitation intensities ranged from 0.15~$\mathrm{Wcm^{-2}}$ to 7.90~$\mathrm{Wcm^{-2}}$ for sample S1 and from 0.15~$\mathrm{Wcm^{-2}}$ to 3.95~$\mathrm{Wcm^{-2}}$ for sample S2. After being dispersed by a grating monochromator, the PL light was detected using an InGaAs line detector at the temperature of $-100\,\mathrm{^{\circ}C}$.

\section{Results and discussion} \label{res}
\subsection{Analysis of the photoluminescence spectra \label{PLspectra}}

\begin{figure}[ht]
\includegraphics{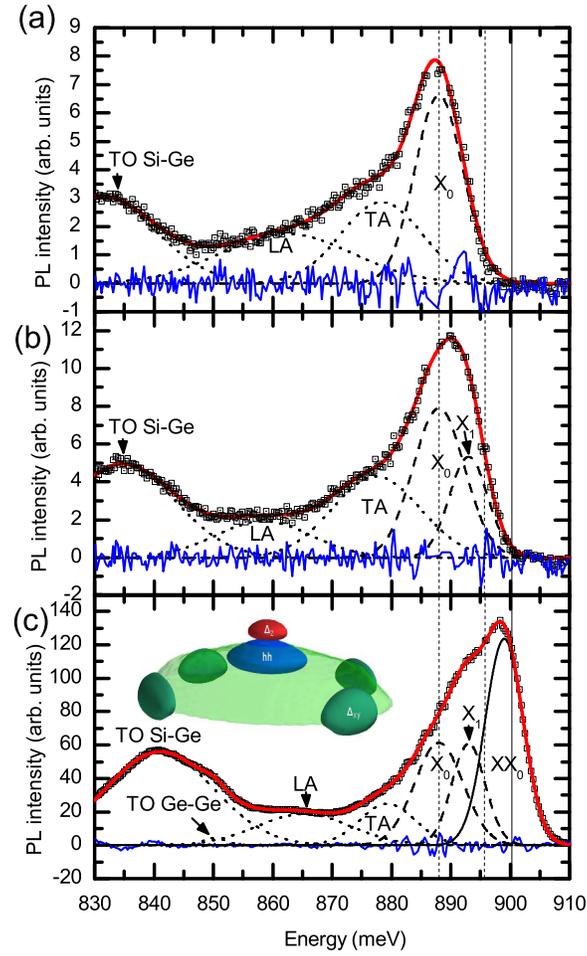}
\caption{
Measured (open squares) and fitted (red curve) PL spectrum of the sample S1 for (a) $I=0.25\,\mathrm{Wcm^{-2}}$, (b) $I=0.49\,\mathrm{Wcm^{-2}}$ and (c) $I=4.94\,\mathrm{Wcm^{-2}}$. The residual error (blue curve) has been multiplied by a factor of 2. The individual GL profiles are attributed to phonon replicas (dotted curves), excitonic ($X_0,\,X_1$, broken curves) and biexcitonic ($XX_0$, full curve) states. The calculated values of the energies of the excitonic (biexcitonic) states for the model dome-shaped dot (see Fig.~\ref{figSt}) blueshifted by 1~meV are displayed by dashed (solid) vertical lines. The inset of panel (c) shows the calculated probability densities (contours of $\mathrm{|\Psi ^2|} = 0.1\,\mathrm{nm^{-3}}$) of the hole ground~state (blue), $\Delta _{xy}$ (green) and $\Delta _{z}$ (red) electron ground~states and their location within the dot (its surface is represented by the light green surface) as obtained by the nextnano++ simulation suite.~\cite{next} 
\label{fig1}}
\end{figure}

\begin{figure}[ht]
\begin{tabular}{c}
\includegraphics{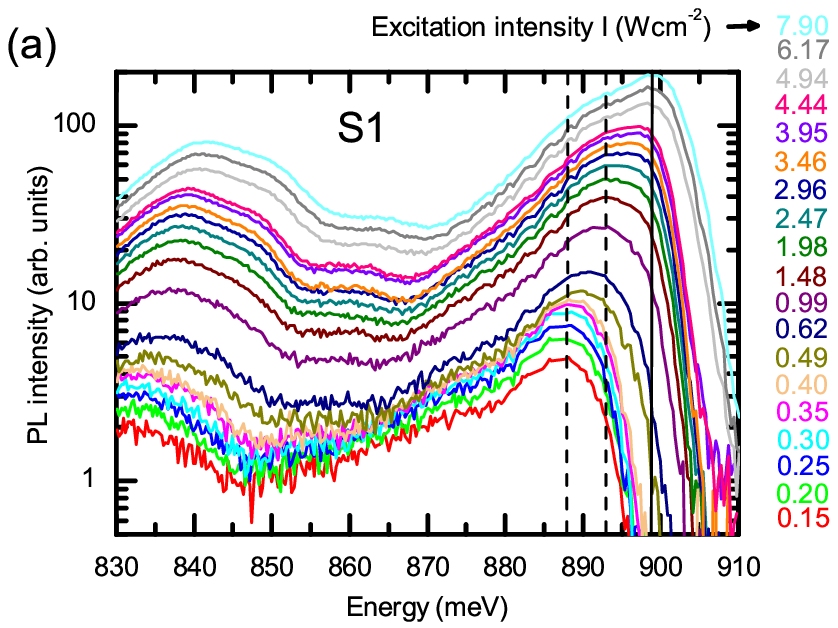}\\
\includegraphics{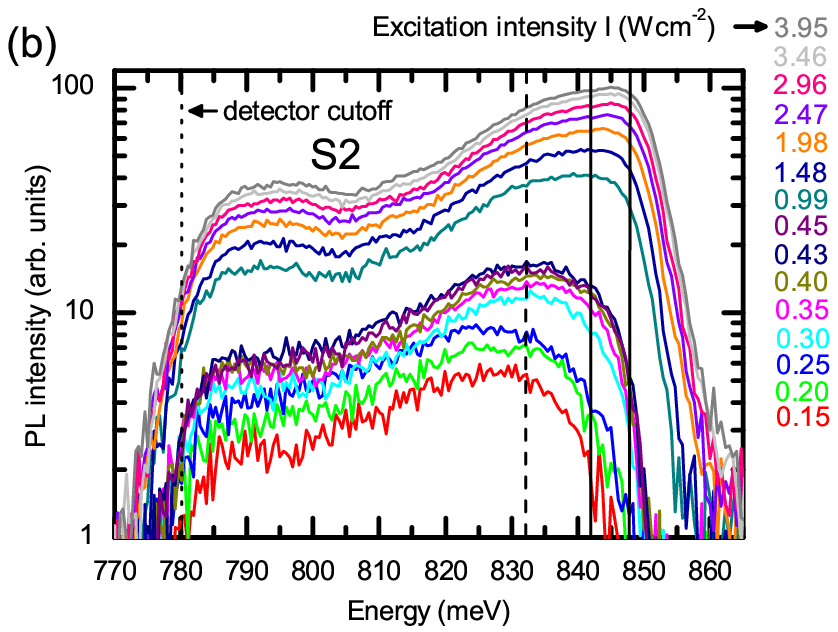}
\end{tabular}
\caption{
Measured PL spectra (the values of the pumping intensity are given next to the right vertical axis) for samples (a) S1 and (b) S2. Dashed (solid) vertical lines correspond to the resonance energies $E_0$ of the fitted GL profiles attributed to excitons (biexcitons). The dotted vertical line in b) corresponds to the detector cutoff energy of 780~meV.
\label{fig2}}
\end{figure}

In Fig.~\ref{fig1}(a)--(c) the PL spectra of S1 are shown for three excitation intensities ($I=$~0.25, 0.49, 4.94\,Wcm$^{-2}$) together with their decomposition into various lines according to the fitting procedure described below. It is evident that with increasing $I$ additional lines appear in the high-energy shoulder of the PL spectra. The observed behavior can not be described by a continuous shift of the emission energy caused by band-bending as a consequence of electron-hole separation due to the type-II band alignment in the SiGe QDs. 
For the complete range of excitation intensities the  PL spectra of S1 and S2 are shown on a logarithmic scale in Figure~\ref{fig2}. Due to the lower growth temperature for the islands in  sample S2 as compared to S1, both the average and the maximum Ge content in S2 is larger.~\cite{BrehmNRL} Thus, the island related PL spectrum of S2 is observed at lower energies than for S1. 
For both samples, the appearance of additional lines with increasing $I$ (as shown in detail in Fig.~\ref{fig1}) is evident over an excitation intensity range between 0.15 and 7.9 (3.95)\,Wcm$^{-2}$ for S1 (S2). 
In the following, a quantitative description of our observations in terms of excitonic contributions will be given.

In order to identify the transitions contributing to the individual PL spectra as a function of $I$, the spectra were fitted using the Gauss-Lorentz (GL) profiles, employing the method of the rational approximants.~\cite{Hum} 
Three parameters were used to describe each contribution: the resonant energies ($E_{0,i}$), the oscillator strengths ($F_i$) and the Gaussian widths ($\Gamma_{G,i}$), where $i$ indexes the GL profiles. For the spectral region of phonon replicas below $885\,\mathrm{meV}$ ($828\,\mathrm{meV}$) for S1 (S2) (see also Fig.~\ref{fig1}), the values of all these parameters were adjusted to fit the data, while for the region of no-phonon transitions the values of $E_{0,i}$ were fixed and only those of $F_i$ and $\Gamma_{G,i}$ allowed to vary in the fitting routine. 
The fixing of the values of $E_{0,i}$ has been motivated by our experimental finding that in the no-phonon region the lines indeed appear at $E_{0,i}$ 
and are fitted more precisely with increasing $I$. This is in agreement with the assumption that 
every profile in the region of no-phonon transitions corresponds to one excitonic or biexcitonic transition. 
The width of the Lorentzian contributions $\Gamma_{L,i}$ to the fitted line width was fixed at the small value of 0.001 meV for all the fitted GL profiles, i.e., the profiles were treated as almost purely Gaussian ones. The Lorentzian width is negligible at low temperatures because the dominant spectral broadening is inhomogeneous, originating mainly in the statistical variation of the QD structure. 

We have fitted every PL spectrum using the smallest number of profiles needed for a reasonably good fit, assessed by the residual sum of squares and the correlation coefficients between the fitted parameters. The values of the parameters obtained for $I_n$ were used as starting values for fitting the spectrum measured at $I_{n+1}$. Here $I_n$ and $I_{n+1}$ denote subsequent excitation intensities in the series shown in Fig.~\ref{fig2}. 
If, for the given number of profiles the best agreement at $I_{n+1}$ was significantly worse than that at $I_n$, an additional profile was added. In this manner we have identified, for both samples, the number of profiles 
of the phonon assisted part of the spectra to be 3 or 4, and the number of the no-phonon part ranging from 1 to 3. 
As an example, we show in Fig.~\ref{fig1} three selected PL spectra of the sample S1 along with the decomposition into the individual GL profiles. 
The values of $E_{0,i}$ are summarized in Tab.~\ref{t1}. 

The phonon assisted transitions were attributed to the various SiGe phonon modes typically observed in the PL spectra of bulk SiGe samples~\cite{Weber} and quantum wells.~\cite{Sturm} 
With increasing value of $E_0$ these are the Si-Ge transverse optical (TO), the Ge-Ge TO (for S1 resolved only at $I>1.48\,\mathrm{Wcm^{-1}}$), the longitudinal acoustic (LA) and the transverse acoustic (TA) (for S2 resolved only at $I>0.20\,\mathrm{Wcm^{-1}}$) phonon assisted transitions. The Si-Si TO phonon assisted transition was not identified in our spectra; its contribution may overlap with the band attributed to the Si-Ge TO phonon replica. With respect to the no-phonon transitions, for the sample S1 (S2) and for $I$ in the range from 0.15 to 0.40~$\mathrm{Wcm^{-2}}$ (0.15 to 0.30~$\mathrm{Wcm^{-2}}$) a single profile with a fixed value of $E_0=888\,\mathrm{meV}$ ($E_0=832\,\mathrm{meV}$) was used. For $I$ in the range from 0.49 to 0.62~$\mathrm{Wcm^{-2}}$ (0.35 to 0.99~$\mathrm{Wcm^{-2}}$), a second profile with $E_0=893\,\mathrm{meV}$ ($E_0=842\,\mathrm{meV}$) was added and another profile with $E_0=898\,\mathrm{meV}$ ($E_0=848\,\mathrm{meV}$) was added for $I$ in the range from 0.99 to 7.90~$\mathrm{Wcm^{-2}}$ (1.48 to 3.95~$\mathrm{Wcm^{-2}}$). The purely electronic (no-phonon) transitions are attributed to the ground state exciton ($\mathrm{X_0}$), excited exciton ($\mathrm{X_1}$), and ground state biexciton ($\mathrm{XX_0}$) in the case of S1, see Fig.~\ref{fig1}, and to $\mathrm{X_0}$, $\mathrm{XX_0}$ and excited biexciton state ($\mathrm{XX_1}$) in the case of S2. This interpretation is based on the observed $I$ dependence of the oscillator strengths $F_i$ (see subsection \ref{Idep}), and the results of electronic structure calculations (see subsection \ref{elcalc}).

\subsection{Excitation intensity dependence \label{Idep}}

\begin{figure}[ht]
\begin{tabular}{c}
\includegraphics{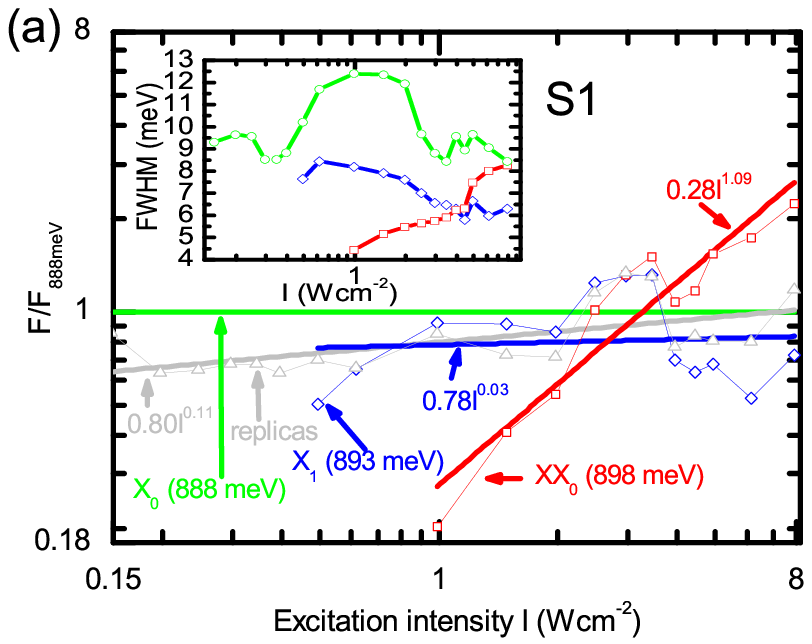}\\
\includegraphics{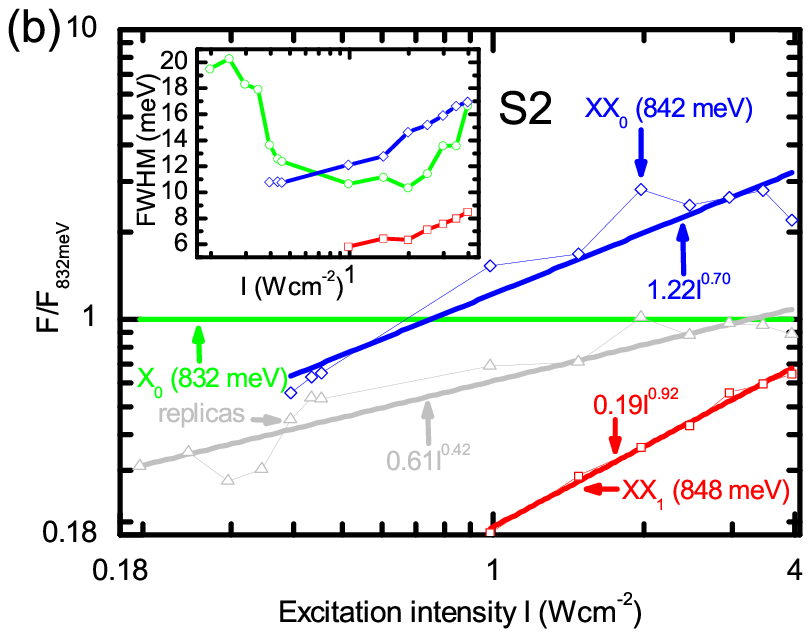}
\end{tabular}
\caption{Excitation intensity ($I$) dependence of the oscillator strength ($F_{i}$) normalized to the respective squared resonance energies $E_{0,i}$ and divided by the oscillator strength $F_{X_0}$ of the no-phonon transition with the lowest value of $E_0$, i.e. of the $X_0$ profile, for the no-phonon transition GL profiles in the sample S1 (a) and in the sample S2 (b). The sums of $F_i$ of the phonon replicas are displayed by open triangles. 
The thick lines represent fits of the measured data to the linear functions $a_1\cdot I^{a_2}$ and the thin ones are guides to the eye. Note, that due to the linearity of the normalization factor $F_{X_0}$ with respect to $I$, quadratic (linear) powers of $I$ appear as linear (constant) functions. The graphs in the insets show the $I$ dependence of the FWHM of the fitted Gaussian lines. The same symbols as in the main diagram are used except for the dependencies of the profiles with the lowest value of $E_0$ which are displayed by open circles.  
\label{fig3}}
\end{figure}

The $I$ dependencies of the oscillator strengths $F_{i}$ and of the Gaussian full widths at half maximum (FWHM) of the fitted profiles of the no-phonon transitions are displayed 
in Fig.~\ref{fig3} and its insets, respectively. In order to suppress the uncertainties of the 
pumping intensities all $F_i$ values shown in Fig.~\ref{fig3} for sample S1 (shown in panel a)  [S2 (panel b)] are normalized to the oscillator strength 
of the 888\,meV (832\,meV) band, which 
corresponds to the lowest excitonic state ($X_0$, with $F_{X_0}$ linear in $I$).  Figure~\ref{fig3}(a) reveals an approximately linear ($\sim I^{1.03}$) dependence of $F$ of the $893\,\mathrm{meV}$ band and a superlinear ($\sim I^{2.09}$) dependence of the $898\,\mathrm{meV}$ band of the sample S1. For the sample S2, both the $842\,\mathrm{meV}$ band and the $848\,\mathrm{meV}$ one exhibit a superlinear, ($\sim I^{1.70}$) and ($\sim I^{1.92}$) dependence of $F$, respectively, as shown in Fig.\,\ref{fig3}(b) (Note, that due to the linearity of the normalization factor $F_{X_0}$ with respect to $I$, quadratic (linear) powers of $I$ appear as linear (constant) functions in Fig.~\ref{fig3}).

Figure~\ref{fig3} also shows the $I$ dependence of the sum of the oscillator strengths of the phonon replicas. This dependence is also slightly superlinear, $\sim I^{1.11}$ and $\sim I^{1.42}$ for S1 and S2, respectively. We interpret this finding as an indication for contributions of phonon replica lines of non-linear exciton emissions. Note, that for the sample S2 the relative magnitude of the contribution to the phonon-replica bands superlinear in $I$ 
is larger than for the sample S1. This may be due to the fact that the ratio of the sum of $F$ for the no-phonon bands having a superlinear $I$ dependence to that of those having a linear $I$ dependence is higher for S2 than for S1. 
The broad features in the $I$ dependence of $F$ and FWHM (the latter shown in the inset of Fig.~\ref{fig3}) are fitting artifacts caused by high correlations of the parameters $F_i$ and $\Gamma_{G,i}$, respectively. Note, that the values of the FWHM of the no-phonon bands are comparable to the FWHM of the $\mathrm{X_0}$ band observed for SiGe bulk crystals (8~meV, see Fig.~10 of Ref.~\onlinecite{Weber}). However, they are considerably larger for S2 than for S1 (by a factor of $\sim 2$). 

The insets of Fig.~\ref{fig3} show that the differences between the various exciton resonance energies are comparable to the widths of the GL profiles, i.e. despite the excellent homogeneity of the QD ensemble these differences are just beyond the experimental resolution limit. Thus, any additional inhomogeneous line broadening, such as that occurring in randomly nucleated islands, hindered the discrimination of various excitonic contributions to the PL spectra in previous studies. 


\subsection{Electronic structure calculations\label{elcalc}}

To get a better insight into the origin of the no-phonon transitions, a series of calculations of the electronic structure was performed using the following two-step approach. First, the single particle wavefunctions were obtained by the nextnano++ solver~\cite{next} using the single band effective Schr\"{o}dinger equation for the electron states in the $\Delta$ valley of the lowest conduction band and the six-band envelope function method 
for the  hole states. The electron and hole states were thus treated as decoupled. This approximation is justified considering the energy difference of $\sim 750$~meV between the extrema of the confinement potentials of electrons and holes. 
Second, these calculated wavefunctions were used to construct a basis set for the configuration interaction (CI) calculations.~\cite{Rodt} 
The evaluation of the 
Coulomb matrix elements is facilitated by the orthogonality of the 
periodic parts of the Bloch waves of the bottom of the conduction band in Si
and the top of the valence band in both Si and Ge.~\cite{Cardona} 

\begin{figure}[ht]
\renewcommand{\tabcolsep}{2pt}
\begin{tabular}{c}
\includegraphics[width=0.4\textwidth]{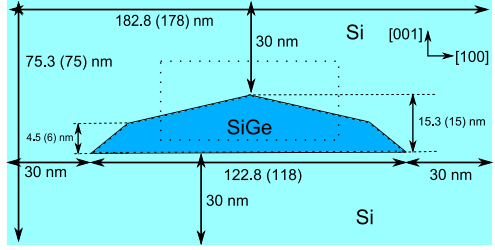} \\
\end{tabular}
\caption{
A cut through the simulation space for the SiGe QD of sample S1 (S2), containing its vertical axis. 
The QDs were approximated by cylindrically symmetric cones with the dimensions shown in the figure. The Ge content inside the QD varies linearly from 0.277 (0.34) at the base of the cone to 0.43 (0.49) at its apex.
The dotted rectangle denotes the space where the Schr\"{o}dinger equation was solved and the grid spacing was set to 0.5~nm.
\label{figSt}}
\end{figure}

Next we describe the model structures. The SiGe QDs of sample S1 were defined on the rectangular grid and approximated by cones with base diameter of 122.8~nm and a height of 15.3~nm as sketched in Fig.~\ref{figSt}. For sample S2, a similar structure was used with slightly different dimensions of 118~nm and 15~nm deduced from the AFM measurements on uncapped islands. In the model structure representing the sample S1 (S2) the Ge content linearly increases from 0.277 (0.34) at the QD base to 0.43 (0.49) at its apex; this is motivated by results of Ref.~\onlinecite{BrehmNRL}. 
The Ge concentration profiles were chosen to 
warrant agreement between the measured and calculated values of the transition energy of the lowest no-phonon transition. For more information on the model structures, see Fig.~\ref{figSt}, for the material parameters, see Tabs.~\ref{ts1},~\ref{ts2}. We have found that the uncertainty in the values of the deformation potentials in Si and Ge (estimated from the spread of the values published in Refs.~\onlinecite{Walle1, Walle2, Cargill, zunger}) leads to an uncertainty of $\sim 20\,\mathrm{meV}$ in the calculated value of the energy of $\mathrm{X_0}$. However, the differences between the energy of $\mathrm{X_0}$ and those of the other excitonic complexes are almost independent of the choice of the deformation potential parameters. The spacing of the grid used in the calculations was set to 4~nm in both lateral and vertical directions except for a cuboid around the QD apex (shown in Fig.~\ref{figSt} by the broken line) where the spacing was 0.5~nm in all directions. The Schr\"{o}dinger equation was solved only in this subspace owing to the expected positions of the electron and hole states.~\cite{BrehmNJP} On the other hand, the minimization of the strain energy was performed in the whole simulation space. For both calculations von Neumann boundary conditions were used in both lateral and vertical directions.

\subsection{Assignment of the PL bands\label{bandassign}}

The calculated values of the transition energies of the states $\mathrm{X_0}$, $\mathrm{X_1}$ and $\mathrm{XX_0}$ for the sample S1 are indicated in Fig.~\ref{fig1} by the broken and full vertical lines along with the probability densities of the single electron wavefunctions [inset of panel c)]. 
Note that for both samples the hole wavefunctions were predominantly composed of heavy holes (96\%); the electron wavefunctions, from which the lowest excitonic (and biexcitonic) states were composed, belonged purely to the lower lying $\Delta_z$ conduction band valleys oriented with their main axis along the [001] growth direction. This is due to the difference of about $18\,\mathrm{meV}$ between the energies of the lowest $\Delta_{z}$ and the four $\Delta_{xy}$ single electron states and almost no spatial overlap between the $\Delta_{xy}$ state and the hole states (these results have been obtained by calculations involving the whole simulation space with a less dense grid).~\cite{BrehmNJP} 

The resulting energy of the excitonic ground state ($\mathrm{X_0}$) for the sample S1, calculated using the CI method with 6 electron and 4 hole basis states, was found to be 887~meV (the corresponding single particle energy difference between the electron and hole states is 896~meV), close to the observed value of 888~meV, and that of the first exited excitonic state ($\mathrm{X_1}$) was found to be 894~meV. The biexciton ($\mathrm{XX_0}$) was found to be shifted to higher energies with respect to the exciton ground state by 11~meV, a value which is in very good agreement with the experimentally observed energy difference of 10~meV between the first and the third no-phonon GL profile having the linear and a superlinear $I$~dependence of $F$, respectively. Also the observed difference of $5\,\mathrm{meV}$ between the energies of the first and the second no-phonon profile is in reasonable agreement with the calculated value of $E(\mathrm{X_1})-E(\mathrm{X_0})$ of $7\,\mathrm{meV}$. Thus, we assign the $888\,\mathrm{meV}$ and $893\,\mathrm{meV}$ profiles to the recombination of $\mathrm{X_0}$ and $\mathrm{X_1}$, respectively, and the $898\,\mathrm{meV}$ profile to the recombination of $\mathrm{XX_0}$. This assignment is corroborated by the observed dependence of the strength of the respective GL profiles on $I$ as discussed in section \ref{Idep}.

For the sample S2, the energy of $\mathrm{X_0}$, calculated using the CI method with 6 electron and 4 hole basis states, was found to be 837~meV (the corresponding single particle energy difference between the electron and hole states is 847~meV), close to the observed value of 832~meV. The calculated values of the blueshift of the biexciton ground state $\mathrm{XX_0}$ and of the biexciton excited state $\mathrm{XX_1}$ are 12~meV and 18~meV, respectively, in good agreement with the experimentally observed energy differences of 10~meV and 16~meV between the second and the first no-phonon band, and between the third and the first no-phonon band, respectively. Thus, we attribute the 832~meV, 842~meV and 848~meV bands to the recombination of $\mathrm{X_0}$, $\mathrm{XX_0}$ and $\mathrm{XX_1}$, respectively, again in agreement with the observed dependence of their strength on $I$ as discussed in section \ref{Idep}. 
The excited excitonic state $\mathrm{X_1}$ observed in sample S1 was not identified here. This is probably due to the larger inhomogeneous broadening of the GL profiles in S2 compared to S1. The $\mathrm{X_1}$ band might contribute to the second no-phonon profile, causing a slightly lower magnitude of its superlinear $I$ dependence seen in Fig.~\ref{fig3}(b). Also, excited state surface trapping, as observed in Ref.~\onlinecite{KambhampatiTJoPCC2011} for colloidal nanocrystals, might quench more efficiently the $\mathrm{X_1}$ emission in S2 as compared to S1 due to the thinner capping layer of S2. The assignment of the 842~meV and 848~meV bands to $\mathrm{XX_0}$ and $\mathrm{XX_1}$ is not unique as a similarly satisfying agreement with the experimental data may be achieved by attributing these to the recombination of the groundstate trion and a $\mathrm{XX_0}$ at an even higher energy, respectively. However, in our calculations this would require the assumption of a Ge concentration profile in the QD, with an -- for the ordered SiGe QDs -- unrealistically large Ge accumulation of up to 60-70\% at its apex. Evidence for such a large Ge concentration in the apex of SiGe transition dome and dome islands have been found up to now only in randomly nucleated islands grown on planar substrates, but not in ordered ones.~\cite{Rastelli,Schulli,BrehmNRL} 

Our calculations and experiments show that the exciton-exciton interaction is pronouncedly anti-binding in SiGe QDs, resulting in $\sim 10\,\mathrm{meV}$ higher biexciton transition energies as compared to the exciton ones. A similar antibinding exciton-exciton interaction was invoked in the interpretation of Ge hut-cluster absorption spectra.~\cite{Yakimov1} Since also in InAs/GaAsSb\cite{Kle}, GaSb/GaAs \cite{MatsudaAPL2007} and in InP/GaAs \cite{BansalPRB2009} quantum dot systems with type-II band alignment a large anti-binding exciton-exciton interaction was found, we conclude that the antibinding character of the exciton-exciton interaction is characteristic of type-II systems with spatially separated electrons and holes.~\cite{Kle,YoffeRev}. Our conclusion is also supported by extensive work on exciton-exciton repulsion in colloidal semiconductor nanocrystals reported in Ref.\,\onlinecite{KlimovN2007,DeutschPCCP2011}.   

\section{Conclusions\label{conc}}

In conclusion, we have performed extensive intensity dependent PL measurements on two ensembles of two dimensionally ordered SiGe QDs, whose excellent homogeneity is due to the controlled growth on prepatterned Si substrates. The spectra were decomposed into a series of distinct bands with characteristic excitation intensity dependencies of their oscillator strengths. Electronic structure calculations were performed using the nextnano++ solver and the calculated wavefunctions were used as a basis set for configuration interaction calculations. Based on these calculations, the transition energies of the $\mathrm{X_0}$,$\mathrm{X_1}$,$\mathrm{XX_0}$ and $\mathrm{XX_1}$ states were compared with the experimentally observed PL bands. Together with the experimentally observed excitation intensity dependence of the various PL bands, the excitonic and biexcitonic recombinations are identified in this type-II quantum dot system. To the best of our knowledge this is the first clear evidence for the formation of excitonic complexes in this system, favored by the zero-dimensional nature of the SiGe/Si QDs. The determined exciton interaction and excited state energies are important for the application of SiGe QDs in Si photonics.

\section{Acknowledgements\label{ack}}

The authors thank M. Grydlik for her help with designing the growth of the ordered QDs. P.~K. would like to thank the \"OAD Stipendium ''Aktion \"Osterreich--Tschechien'' and the NANOE project No. CZ.1.07/2.3.00/20.0027 for financing his research stay in Linz. This work was supported by the project "CEITEC - Central European Institute of Technology" (CZ.1.05/1.1.00/02.0068) from European Regional Development Fund, the internal project MUNI/A/1047/2009, the Austrian Nanoinitiative (FFG, Grant Nos. 815802 and 815803), the IRON Project Nos. SFB025-02 and SFB025-12 (FWF) and GMe, Austria.


%

 \newpage

\clearpage

\begin{widetext}
\begin{table*}[ht]
\caption{
Energies of the fitted GL profiles and calculated energies of the no-phonon excitons for samples S1 and S2 along with the energies of the phonon replicas taken from Fig.~8 of Ref.~\onlinecite{Weber}. The energies of the phonon replicas and the excitonic complexes are given relative to the corresponding value of $\mathrm{X_0}$. The displayed values of the energies of the phonon replicas were taken from the corresponding GL profiles of the spectra for the lowest excitation intensities. The estimated uncertainties of the energies of the fitted GL bands are $\sim 3$ meV for the no-phonon bands and $\sim 6$ meV for the phonon assisted ones. For the uncertainties of the calculated values of $\mathrm{X_0}$ see text, the uncertainties of the calculated energies of the other excitonic complexes is $\sim 2$ meV. The uncertainty of the phonon energies derived from Fig.~8 of Ref.~\onlinecite{Weber} is $\sim 1$ meV. 
\label{t1}
}
\begin{tabular}{|l|c|ccc|cccc|}
\hline
&$\mathrm{X_0}$&$\mathrm{X_1}$& $\mathrm{XX_0}$& $\mathrm{XX_1}$& TO Si-Ge& TO Ge-Ge& LA& TA\\
\hline
S1 fit (meV)& $888 \pm 3$& $+5 \pm 3$& $+10 \pm 3$& -& $-52 \pm 6$&$-39 \pm 6$&$-26 \pm 6$&$-9 \pm 6$\\
S1 theory (meV)& $887\pm 20$& $+7\pm 2$& $+11\pm 2$& -& -& -& -& -\\
\hline
S2 fit (meV)& $832 \pm 3$& -& $+10 \pm 3$& $+16 \pm 3$& $-46 \pm 6$&$-34 \pm 6$&$-19 \pm 6$&$-9 \pm 6$\\
S2 theory (meV)& $837\pm 20$& $+8\pm 2$ & $+12\pm 2$& $+18\pm 2$& -& -& -& -\\
\hline
Ref.~\onlinecite{Weber} (meV)& -& -& -& -& $-49$& $-35$& $-30$& $-10$\\
\hline
\end{tabular}

\end{table*}
\end{widetext}

\begin{widetext}
\begin{table*}[ht]
\caption{Description of the material parameters used in the calculations, $\gamma_1$, $\gamma_2$ and $\gamma_3$ are the Luttinger parameters.~\cite{Lutt} \label{ts1}}

\begin{tabular}{lc}
\hline \hline
Parameter & Description\\
\hline
$a$ & lattice constant\\
$a_{exp}$ & lattice thermal expansion coefficient\\
$C_{11}$ & elastic constant\\
$C_{12}$ & elastic constant\\
$C_{44}$ & elastic constant\\
$\varepsilon_{r}$ & static dielectric constant\\
$m_{\Delta l}$ & $\Delta$-valley longitudinal electron effective mass\\
$m_{\Delta t}$ & $\Delta$-valley transversal electron effective mass\\
$E_0$ & bandgap\\
$\alpha$ & Varshni parameter\\
$\beta$ & Varshni parameter\\
$E_v$ & valence band offset\\
$\Delta_0$ & spin-orbit split-off energy\\
$a_c$ & absolute deformation potential for conduction band $\Delta$-valley\\
$a_u$ & uniaxial shear deformation potential of the conduction band $\Delta$-valley\\
$a_v$ & absolute deformation potential for valence band\\
$a_{ub}$ & uniaxial shear deformation potential b of the valence bands\\
$a_{ud}$ & uniaxial shear deformation potential d of the valence bands\\
$L$ & Dresselhaus parameter;~\cite{Dress} $L=-\gamma_1-4\gamma_2-1$\\
$M$ & Dresselhaus parameter;~\cite{Dress} $M=2\gamma_2-\gamma_1-1$\\
$N$ & Dresselhaus parameter;~\cite{Dress} $N=-6\gamma_3$\\
\hline \hline
\end{tabular}
\end{table*}

\begin{table*}[ht]
\caption{Values of the material parameters used in the calculations. The unit $m_0$ represents the free electron mass. \label{ts2}} 

\begin{tabular}{llccc}
\hline \hline
Parameter & Unit & Si & Ge & $\mathrm{Si_{1-x}Ge_x}$ \\
\hline
$a$ & \AA & 5.4304~[\onlinecite{landoltbornstein}] & 5.6579~[\onlinecite{landoltbornstein}] & linear \\
$a_{exp}$ & \AA /K& 1.8138$\times$10$^{-5}$~[\onlinecite{landoltbornstein}] & 5.8$\times$10$^{-5}$~[\onlinecite{sze}] & linear\\
$C_{11}$ & GPa& 165.77~[\onlinecite{landoltbornstein}] & 128.53~[\onlinecite{landoltbornstein}] & linear\\
$C_{12}$ & GPa& 63.93~[\onlinecite{landoltbornstein}] & 48.26~[\onlinecite{landoltbornstein}] & linear\\
$C_{44}$ & GPa& 79.62~[\onlinecite{landoltbornstein}] & 66.80~[\onlinecite{landoltbornstein}] & linear\\
$\varepsilon_{r}$ & -& 11.7~[\onlinecite{boer}] & 16.2~[\onlinecite{landoltbornstein}] & linear\\
$m_{\Delta l}$ & $m_0$ & 0.916~[\onlinecite{boer}] & 1.350~[\onlinecite{next}] & linear\\
$m_{\Delta t}$ & $m_0$ & 0.190~[\onlinecite{boer}] & 0.290~[\onlinecite{next}] & linear\\
$E_0$ & eV & 1.17~[\onlinecite{ioffe}] & 0.931~[\onlinecite{Weber}] & 0.931x + 1.17(1-x) - 0.206x(1-x)~[\onlinecite{Weber}]\\
$\alpha$ & eV/K & 0.473$\times$10$^{-3}$~[\onlinecite{sze}] & 0.4774$\times$10$^{-3}$~[\onlinecite{next}] & linear\\
$\beta$ & K & 636~[\onlinecite{sze}] & 235~[\onlinecite{next}] & linear\\
$E_v$ & eV & 1.090~[\onlinecite{zunger}] & 1.67~[\onlinecite{zunger}] & linear\\
$\Delta_0$ & eV & 0.044~[\onlinecite{landoltbornstein}] & 0.30~[\onlinecite{landoltbornstein}] & linear\\
$a_c$ & eV & 3.40~[\onlinecite{zunger}] & 0.14~[\onlinecite{zunger}] & linear\\
$a_u$ & eV & 9.16~[\onlinecite{Walle2}] & 9.42~[\onlinecite{Walle2}] & linear\\
$a_v$ & eV & 2.05~[\onlinecite{zunger}] & -0.35~[\onlinecite{zunger}] & linear\\
$a_{ub}$ & eV & -2.10~[\onlinecite{laude}]& -2.86~[\onlinecite{chandra}]& linear\\
$a_{ud}$ & eV & -4.85~[\onlinecite{laude}]& -5.28~[\onlinecite{chandra}]& linear\\
$L$ & $\hbar^2/2m_0$ & -6.69~[\onlinecite{rieger}] & -31.34~[\onlinecite{hensel}] & linear\\
$M$ & $\hbar^2/2m_0$ & -4.62~[\onlinecite{rieger}] & -5.90~[\onlinecite{hensel}] & linear\\
$N$ & $\hbar^2/2m_0$ & -8.56~[\onlinecite{rieger}] & -34.14~[\onlinecite{hensel}] & linear\\
\hline \hline
\end{tabular}
\end{table*}
\end{widetext}

\end{document}